\documentclass[runningheads]{rnl_modified}

\usepackage[francais, english]{babel}
\usepackage[utf8]{inputenc} 			
\usepackage[T1]{fontenc}				
\usepackage{lmodern}					
\usepackage{cite}
\usepackage{array}
\usepackage{amsmath,amsfonts,amssymb}
\usepackage{graphicx}
\usepackage{color}

\begin{document}
\title*{Locating a regular needle in a chaotic haystack, and conversely, using Lyapunov Weighted Dynamics}

\titrecourt{Lyapunov Weighted Dynamics}

\author{Tanguy Laffargue
\and Julien Tailleur
}

\index{Laffargue Tanguy}              
\index{Tailleur Julien}
 
\adresse{Laboratoire Matière et Systèmes Complexes (CNRS UMR 7057), Université Paris Diderot, 10 rue Alice Domon et Léonie Duquet, 75205 Paris Cedex 13, France}

\email{tanguy.laffargue@univ-paris-diderot.fr}

\maketitle   

\begin{resume}
	\foreignlanguage{francais}{La dynamique de certains systèmes physiques est gouvernée par des structures de chaoticité atypique. Ces structures, qui représentent un faible volume dans l'espace des phases, sont difficiles à localiser. Nous présentons dans cet article un algorithme, la dynamique biaisée par les Lyapunov, permettant de les mettre en évidence.} 
\end{resume}

\begin{resumanglais}
	In many physical systems, dynamics is  ruled by structures of atypical chaoticity. These structures may occupy a very small volume in phase space and can thus be very difficult to locate numerically. In this article, we review an algorithm, the Lyapunov Weighted Dynamics, which efficiently reveals trajectories of atypical chaoticity.
\end{resumanglais}

\section{Introduction}
Structures of atypical chaoticity, although rare, can play an important role in many physical systems. For instance, resonances and separatrices play a crucial part in determining stability of planetary systems \cite{LaffargueLaskar, LaffargueMurrayHolman}. Similarly, to study the global diffusion mechanism in almost-integrable systems, we need to focus on extremely thin chaotic layers which are responsible for Arnold diffusion \cite{LaffargueArnold, LaffargueArnoldWeb1, LaffargueArnoldWeb2}. Likewise, unstable objects like solitons and chaotic breathers \cite{LaffargueFPU} are responsible for the energy transport in Bose-Einstein condensates \cite{LaffargueBoseEinstein} and in biological molecules \cite{LaffargueMolecules}. 

Those structures are usually not only rare but also unstable, which makes them even harder to find. Despite the progress made in the last few years, most numerical methods to locate those structures are restricted to low-dimensional systems or are model-specific. The Lyapunov Weighted Dynamics is a Monte Carlo algorithm which samples trajectories according to their Lyapunov spectrum, an observable measuring the sensitivity to initial conditions and hence chaoticity. In this article, we review this algorithm and show how it can be used to reveal rare trajectories, impossible to find with direct simulations, in both low and high dimensions, opening the door to applications going from celestial mechanics to statistical physics.

\section{The Lyapunov spectrum, a large deviation problem}
For a dynamical system defined by trajectories of $D$-dimensional variables $\mathbf{x}(t)$, consider two infinitely close points $\mathbf{x}(0)$ and ${\mathbf{x}(0) + \mathbf{u}(0)}$. The distance $\mathbf{u}(t)$ between them typically grows as 
\begin{equation}
	\left| \mathbf{u}(t) \right| \equiv \left|\mathbf{u}(0)\right| e^{t \lambda_{1}(t)}
\end{equation}
where $\lambda_{1}(t)$ is called the largest finite-time Lyapunov exponent (at time $t$). It measures the sensitivity of the dynamical system to an initial perturbation in the vicinity of $\mathbf{x}(0$). Similarly, we can consider $k+1$ nearby points defining $k$ noncollinear vectors $\mathbf{u}_i(0)$, with ${i \in \{1, \dots, k\}}$, and look at how the area ${V_{k}(t) \equiv \left| \mathbf{u}_{1}(t) \wedge \dots \wedge \mathbf{u}_{k}(t) \right|}$ evolves. In general, for $k \leqslant D$, it grows as 
\begin{equation}
	V_{k}(t) \approx e^{t [\lambda_{1}(t) + \cdots + \lambda_{k}(t)]}
\end{equation}
with ${\lambda_{1}(t) \geqslant \lambda_{2}(t) \geqslant \dots \geqslant \lambda_{k}(t)}$. These are the $k$ largest (finite-time) Lyapunov exponents. Under general assumptions, the $D$ Lyapunov exponents converge as $t$ goes to infinity to finite values, yielding the so-called Lyapunov spectrum. In the following, in order to facilitate the comprehension, we will focus on the largest one, $\lambda_{1}$, that we simply call $\lambda$, but everything said below can be generalized to the entire Lyapunov spectrum. 

The Lyapunov exponent $\lambda(t)$ can fluctuate between two trajectories: it does not take a unique value, and is distributed according to a distribution $P(\lambda, \,t)$, giving the probability density to observe a trajectory $x(t)$ with an exponent $\lambda$. In the large time limit, this pdf typically obeys a large deviation principle \cite{LaffargueTouchette}
\begin{equation}
	P(\lambda, \,t) \underset{t \to +\infty}{\approx}e^{- t s(\lambda)} \quad \text{ with } \quad s(\lambda) \underset{t \to + \infty}{=} {\cal O}(1).
\end{equation}
$P(\lambda, \,t)$ thus becomes sharper and sharper as time increases and concentrates around a typical value $\lambda^{*}$, which satisfies ${s'(\lambda^{*}) = 0}$. This is why direct simulations of long trajectories are not efficient to isolate trajectories with atypical value of $\lambda$: when $\lambda - \lambda^{*} \sim {\cal O}(1)$, then $s(\lambda) \sim {\cal O}(1)$ and one needs an exponentially large number of
independent random samples ($\sim e^{t s(\lambda)}$) to observe with probability one a trajectory with an exponent $\lambda$.

\section{Thermodynamic formalism}
Brute-force sampling imposes flat measure on the trajectory space by giving the same weight to all trajectories. On the contrary, collecting trajectories with a given $\lambda$ resembles the construction of the microcanonical ensemble in equilibrium statistical physics, where one tries to collect all configurations of given energy $E$. This is a notoriously difficult problem; it is usually simpler to fix the mean value of the energy, by introducing a conjugate parameter, the temperature $\beta$: this is the construction of the canonical ensemble. We will follow a similar strategy here: rather than collecting all trajectories of exponent $\lambda$, we introduce a conjugate parameter $\alpha$ and define the canonical weights:  
\begin{equation}
	P_{\alpha}(\lambda, \,t) \equiv \frac{1}{Z(\alpha, \,t)} P(\lambda, \,t) \, e^{\alpha \lambda t} \underset{t \to +\infty}{\approx} e^{t [\alpha \lambda - s(\lambda) - \mu(\alpha)]}
	\label{laffargue_canonical_weight} 
\end{equation}
where ${Z(\alpha, \,t) \equiv \left< e^{\alpha \lambda t} \right>}$ is the (dynamical) partition function (or in, a more mathematical language, the moment-generating function). With those new weights, the new typical Lyapunov exponent $\lambda^{*}_{\alpha}$ satisfies ${s'(\lambda^{*}_{\alpha}) = \alpha}$. The conjugate parameter $\alpha$ acts like a temperature for chaoticity: positive $\alpha$ favors trajectories with large Lyapunov exponents, hence chaos, whereas negative $\alpha$ favors trajectories with small Lyapunov exponents, and thus promotes stability. Furthermore, in the canonical ensemble, all the macroscopic (static) properties can be extracted from the partition function or from the free energy. Here also, we can define a dynamical free energy $\mu(\alpha)$ by
\begin{equation}
	Z(\alpha, \,t) \underset{t \to +\infty}{\approx} e^{t \mu(\alpha)}.
\end{equation}
It relates to the dynamical entropy by a Legendre-Fenchel transform: 
\begin{equation}
	\mu(\alpha) = \underset{\lambda}{\sup} \left[\alpha \lambda - s(\lambda)\right].
\end{equation}
In a more mathematical language, $\mu$ is the cumulant-generating function. The analogy with equilibrium statistical physics is summarized in table \ref{laffargue_thermodynamic_formalism}.

\begin{table}[h]
		\renewcommand{\arraystretch}{2.2}
		\centering
		\begin{tabular}{|>{\centering\arraybackslash}p{3cm}
						|>{\centering\arraybackslash}p{5cm}
						|>{\centering\arraybackslash}p{5cm}|}
			\hline
			\bf Variable & \bf Equilibrium statistical physics & \bf Dynamical system\\
			\hline
			Macrostate & $\rho=\frac{E}{V}$ & $\lambda$\\
			\hline
			Volume & $V$ & $t$\\
			\hline	
			Entropy & $s(\rho) \underset{V \to \infty}{=} \frac{k}{V} \ln \Omega(E, V)$ 
				    & $s(\lambda) \underset{t \to \infty}{=} -\frac{1}{t} \ln P(\lambda, t)$\\[0.5em]
			\hline
			Inverse temperature & $\beta$ & $-\alpha$\\

			\hline	
			Partition function & $Z(\beta, V) = \left< e^{- \beta E} \right>$ 
							   & $Z(\alpha, t) = \left< e^{\alpha \lambda t} \right>$\\
			\hline
			Free energy & $f(\beta) \underset{V \to \infty}{=} - \frac{1}{\beta V} \ln Z(\beta, V)$ 
						& $\mu(\alpha) \underset{t \to \infty}{=} \frac{1}{t} \ln Z(\alpha, t)$\\[0.5em]
    		\hline
		\end{tabular}
		\caption{\label{laffargue_thermodynamic_formalism} Thermodynamic formalism for dynamical systems. Differences in prefactors and signs are due to historical reasons: equilibrium statistical physics was constructed to explain thermodynamics and has to take into account previous definitions (temperature, entropy, free energy) whereas thermodynamic formalism was born in the dynamical system community \cite{LaffargueRuelle, LaffargueGrassbergerThermoForm} and remained closer to the probability theory language.}
\end{table}

\newpage
\section{Lyapunov Weighted Dynamics}
The parameter $\alpha$ has no evident physical meaning, it is thus not obvious how the biased weights (\ref{laffargue_canonical_weight})  can be realized: we do not have thermostat for chaoticity in a lab. Lyapunov Weighted Dynamics (LWD) is a population Monte Carlo algorithm, inspired by Diffusion Monte Carlo algorithm and similar, in spirit, to the "go with the winners" algorithms \cite{LaffargueGrassberger}, which aims at fulfilling this role \cite{LaffargueLWDJulien}. The key idea is to evolve a population of copies of the system, called clones, and to copy and kill them in a controlled way.

We consider $N_{c}$ clones $(\mathbf{x}, \mathbf{u})$ of the dynamical system $\dot{\mathbf{x}}(t) = \mathbf{f}(\mathbf{x}(t))$ and a time increment $\mathrm{d}t$. At every time step $t_{n} = n \,\mathrm{d}t$:
\begin{itemize}
	\item each copy evolves with the dynamics $\dot{\mathbf{x}}=f(\mathbf{x})$ and $\dot{\mathbf{u}} = \frac{\partial \mathbf{f}}{\partial \mathbf{x}} \mathbf{u}$
	\item for each clone $j$, we compute $s_{j}(t) = \frac{|\mathbf{u}(t+dt)|}{|\mathbf{u}(t)|} \simeq e^{\lambda \, \mathrm{d} t}$
	\item each clone $j$ is then replaced, on average, by $s_{j}(t)^{\alpha}$ copies
\end{itemize}
Roughly speaking, at time $t$, one clone has yielded $e^{\alpha \lambda t}$ copies. If the initial population was large enough, the ratio between the total number of clones at time $t$ and the initial number of clone yields
\begin{equation}
	\frac{N_c(t)}{N_c(0)} \simeq \left<{e^{\alpha \lambda t}} \right> \approx e^{t \mu(\alpha)}.
\end{equation}
It thus gives access to the partition function and to the free energy. Two important tricks are used: to maintain the population almost constant, we use $w_{j} = N_{c} \,s_{j}/\sum_{j} s_{j}$ instead of $s_{j}$ for calculating the cloning rate and, to prevent degeneracy of clones and enhance the quality of sampling, a small noise, with appropriate properties (energy conservation, momentum conversation, etc), is added to the dynamics.

A simple way to understand why this algorithm works is to think about it as an evolution problem. Cloning plays the role of reproduction, noise the one of mutation and dependence on $\lambda$ of the cloning rate the one of selection. The convergence of the algorithm is then assured by a sort of ``selection pressure'' and changing $\alpha$ is equivalent to modify the fitness landscape.

This algorithm can be generalized to sample the fluctuations of the $k$ first Lyapunov exponents, by considering one chaotic temperature $\alpha_{i}$ for each Lyapunov exponent $\lambda_{i}$ and using Gram-Schmidt orthonormalization procedure. For technical details and numerical implementations, see \cite{LaffargueLWD}.

\section{Normally hyperbolic invariant manifold}
Normally hyperbolic invariant manifolds (NHIMs) with $p$ unstable directions are manifolds invariant under the dynamics and whose normal directions have the structure of saddles, with exactly $p$ unstable directions. If we consider the LWD with $\alpha=1$, we see that the cloning rate exactly compensates the volume contractions and expansions induced by time evolution for trajectories evacuating from saddles with one unstable direction \cite{LaffargueJulienSUSY}. Then the cloning stabilizes the unstable manifold of NHIMs with one unstable direction and the algorithm populates it uniformly. Similarly, taking $\alpha_{i} = 1$ for $i$ in $\{1,\dots,k\}$ stabilizes the unstable manifold of NHIMs with $k$ unstable directions. 

We can illustrate this property on a simple example: two double well potentials. This system has four degrees  of freedom and the Hamiltonian is given by
\begin{equation}
	H(\boldsymbol{q}, \boldsymbol{p}) = \sum_{i=1,2} \left[\frac{p_{i}^{2}}{2} + \frac{(q_i^{2} - 1)^{2}}{4}\right].
	\label{laffargue_DW_def}
\end{equation}

\begin{figure}[h]
	\centering
	\includegraphics[width=\linewidth]{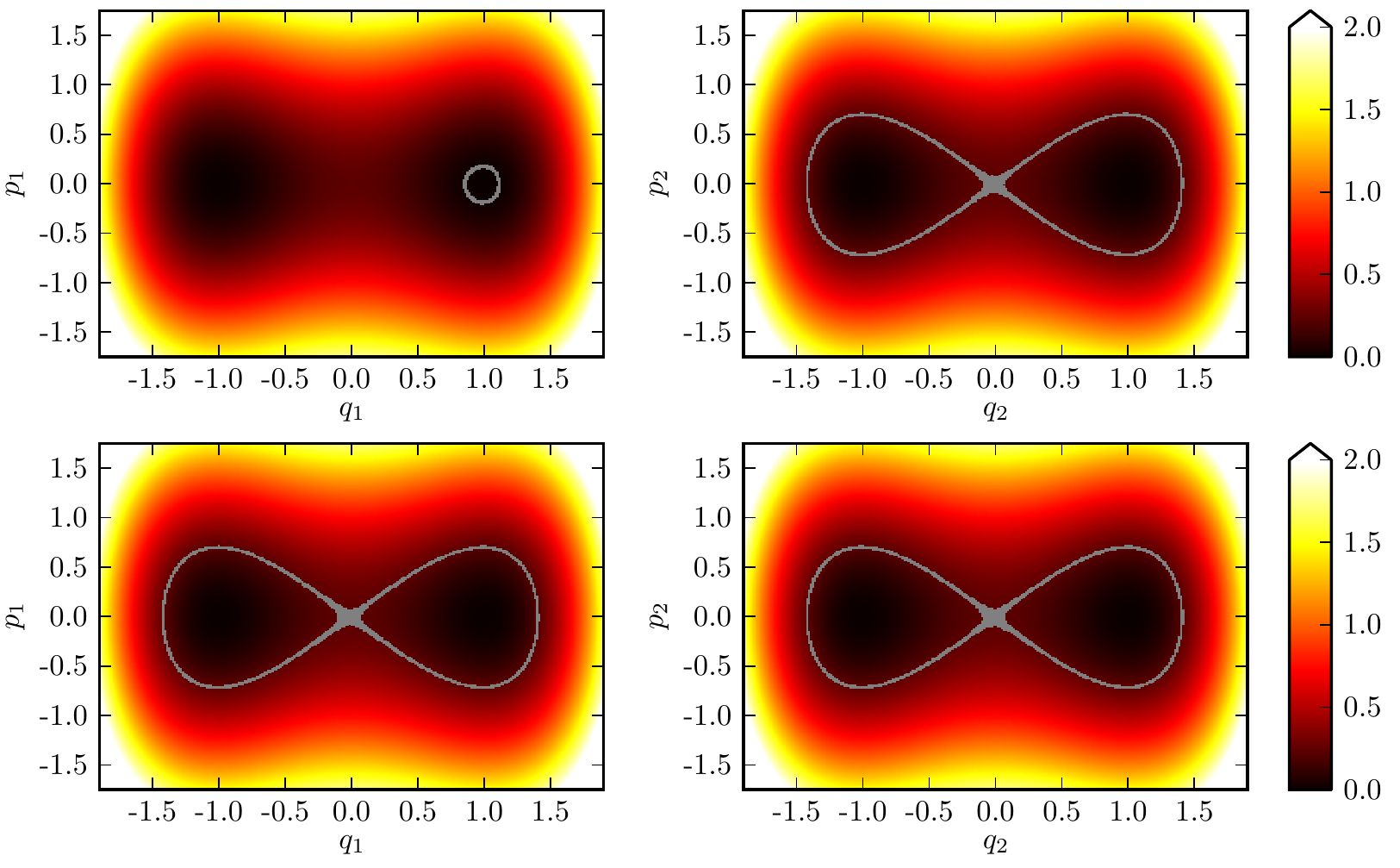}
	\caption{Trajectories of $5\,000$ clones using LWD for the system defined in (\ref{laffargue_DW_def}) for $t \geqslant 250$. The variance of the noise is decreased from $2.10^{-3}$ to $2.10^{-5}$ at $t=60$, and then to $2.10^{-7}$ at $t=120$. The clones are in gray and the color code correponds to the energy $H$. \textbf{Top:} $\alpha_1 = 1$ and $\alpha_{i\geqslant 2} = 0$. \textbf{Bottom:} $\alpha_{1,2} = 1$ and $\alpha_{3,4} = 0$. \label{laffargue_DW}}
\end{figure}

This system has two saddle points, defined respectively by $q_1 = p_1 = 0$ and $q_2 = p_2 = 0$, and, once a Gaussian white noise is added to momenta, its steady-state measure is the flat measure. It has two NHIMs with one unstable direction, corresponding to the Cartesian products between the flat measure over one double well and the saddle point of the other double well. It also has one NHIM with two unstable directions, corresponding to the Cartesian product of the two saddle points. We can see on figure \ref{laffargue_DW} that the LWD with $\alpha_1=1$ isolates the unstable manifold of one NHIM with one unstable direction and that the LWD with $\alpha_{1,2}=1$ isolates the unstable manifold of the NHIM with two unstable directions.

\section{A spatially extended system: the Fermi-Pasta-Ulam-Tsingou chain}
\begin{figure}[h]
	\centering
	\includegraphics[width=\linewidth]{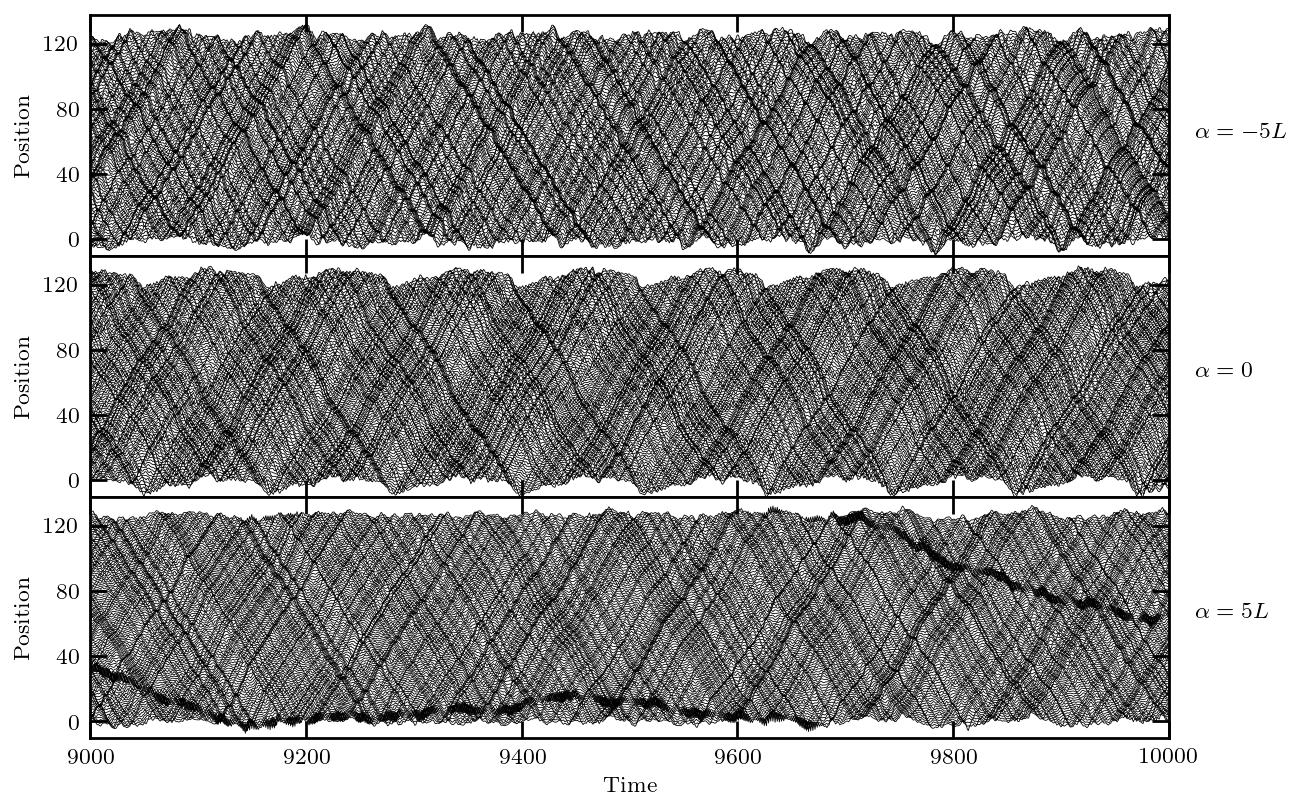}
	\caption{Configuration of one clone of the LWD for the $\beta$-FPU chain with $\beta=0.1$, $L=128$, 200 clones and $H=L$. \textbf{Top:} Gas of solitons. \textbf{Middle:} Equilibrium. \textbf{Bottom:} Chaotic breather.
	\label{laffargue_FPU}}
\end{figure}

This algorithm can be applied to spatially extended systems, like the $\beta$-FPU chain defined by the Hamiltonian
\begin{equation}
	H(\boldsymbol{x}, \boldsymbol{p}) = \sum_{i=1}^{L} \left[ \frac{p_{i}^{2}}{2} + \frac{(x_{i+1} - x_{i})^2}{2} + \beta \frac{(x_{i+1} - x_{i})^4}{4} \right]
\end{equation}	
with periodic boundary conditions $x_{L+1} = x_{1}$. This describes a chain of $L$ particles coupled with anharmonic springs. At equilibrium, the typical configuration is a superposition of short-lived solitons, short-lived chaotic breathers \cite{LaffargueFPU} and thermal fluctuations (phonons). When applying the LWD with $\alpha<0$, we isolate a gas of solitons whereas with $\alpha>0$ we stabilize chaotic breathers. These three cases are illustrated on figure \ref{laffargue_FPU}. In \cite{LaffargueLWDJulien}, fixed boundary conditions were used to isolate solitons, because otherwise the system can put all its energy in a rotation of its center of mass. Here, thanks to a noise which conserves total impulsion, we were able to use periodic boundary conditions.

\begin{figure}[h]
	\centering
	\includegraphics[width=\linewidth]{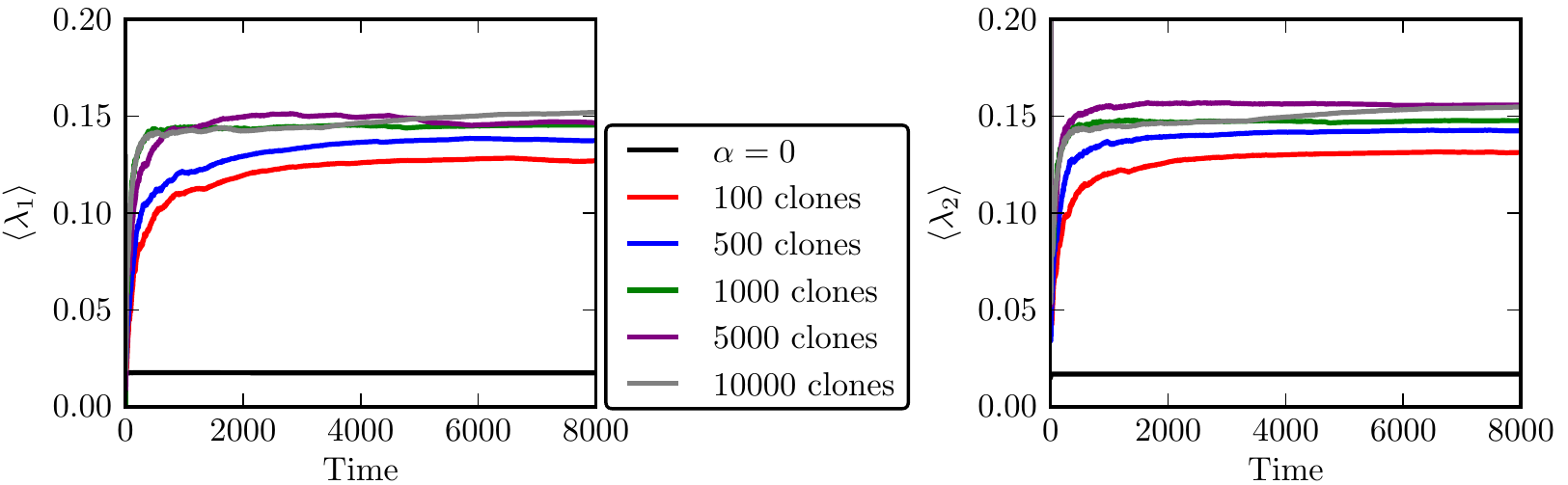}
	\caption{Mean value of the two largest Lyapunov exponents over all the clones and over 5 runs, using LWD for the $\beta$-FPU chain with $\beta=0.1$, $L=200$, $H=L$, $\alpha_{i\neq2} = 0$ and $\alpha_{2} = 5 L$. \label{laffargue_FPU_Lyap}}
	
\end{figure}

Biasing the $k$th Lyapunov exponent with a positive $\alpha_{k}$ reveals really rare trajectories with $k$ non-merging breathers \cite{LaffargueLWD}. We see on figure \ref{laffargue_FPU_Lyap} that, for given $\{\alpha_{i}\}$, the finite-time Lyapunov exponents seem to converge to a finite values $\lambda_{\alpha}$ as time and number of clones increase. This is important because $\lambda_{\alpha}$ can be used to compute the dynamical free energy thanks to thermodynamics integration
\begin{equation}
	\mu(\alpha) = \int_{0}^{\alpha} \lambda_{\alpha'} \, \mathrm{d}\alpha'
\end{equation}
which, compared to direct measurement using $Z(\alpha)$, yields much better (smoother) averages.

\section{Conclusion}
We have seen two applications of this algorithm: the stabilization of the unstable manifold of NHIMs in a simple dynamical system and the detection of localized chaotic breathers in a spatially extended system. In the latter case, we have shown that the measure of the first derivative of the dynamical free energy can be achieved, which opens the way to future studies of dynamical phase transitions in these systems. This algorithm has also been applied elsewhere to localize the Arnold web \cite{LaffargueLWDJulien} and to study the stability of Lagrange points L4 and L5 in the restricted three-body problem \cite{LaffargueLWD}.

\end{document}